\documentclass[pra,amsfonts,amssymb,nofootinbib]{revtex4}
\usepackage{amsmath}
\usepackage{amssymb}
\usepackage{graphics}
\usepackage{latexsym}

\begin{document}
\bibliographystyle{revtex}

\def\Bid{{\mathchoice {\rm {1\mskip-4.5mu l}} {\rm
{1\mskip-4.5mu l}} {\rm {1\mskip-3.8mu l}} {\rm {1\mskip-4.3mu l}}}}

\newcommand{\eL}{{\cal L}}
\newcommand{\half}{\frac{1}{2}}
\newcommand{\J}{\textbf{J}}
\newcommand{\bP}{\textbf{P}}
\newcommand{\G}{\textbf{G}}
\newcommand{\K}{\textbf{K}}
\newcommand{\M}{{\cal M}}
\newcommand{\E}{{\cal E}}
\newcommand{\bu}{\textbf{u}}
\newcommand{\tr}{\mbox{tr}}
\newcommand{\norm}[1]{\left\Vert#1\right\Vert}
\newcommand{\abs}[1]{\left\vert#1\right\vert}
\newcommand{\set}[1]{\left\{#1\right\}}
\newcommand{\ket}[1]{\left\vert#1\right\rangle}
\newcommand{\bra}[1]{\left\langle#1\right\vert}
\newcommand{\ele}[3]{\left\langle#1\left\vert#2\right\vert#3\right\rangle}
\newcommand{\inn}[2]{\left\langle#1\vert#2\right \rangle}
\newcommand{\Real}{\Bid R}
\newcommand{\dmat}[2]{\ket{#1}\!\!\bra{#2}}


\title{A Parametrization of Bipartite Systems Based on $SU(4)$ Euler Angles}

\author{Todd Tilma}
\affiliation{The Ilya Prigogine Center for Studies in Statistical Mechanics
and Complex Systems \\
Physics Department \\
The University of Texas at Austin \\
Austin, Texas 78712-1081}
\email[Email:]{tilma@physics.utexas.edu}
\author{Mark Byrd}
\affiliation{Maxwell Dworkin Laboratory \\
33 Oxford Street \\ 
Harvard University \\
Cambridge, Massachusetts 02138}
\email[Email:]{brd@hrl.harvard.edu}
\author{E.C.G. Sudarshan}
\affiliation{Center for Particle Physics \\
Physics Department \\
The University of Texas at Austin \\
Austin, Texas 78712-1081}
\email[Email:]{sudarshan@physics.utexas.edu}

\date{\today}




\begin{abstract}
In this paper we give an explicit parametrization for all two
qubit density matrices.  This is important for calculations
involving entanglement and many other types of quantum information
processing.  To accomplish this we present a generalized Euler angle parametrization for
$SU(4)$ and all possible two qubit density matrices.  The important 
group-theoretical properties of such a description are then manifest.  
We thus obtain the correct Haar (Hurwitz) measure and volume element for $SU(4)$ which follows from this
parametrization.  In addition, we study the role of this parametrization in
the Peres-Horodecki criteria for separability and its corresponding usefulness
in calculating entangled two qubit states as represented through the parametrization.
\end{abstract}
\maketitle



\pagebreak


\section{Introduction}

In quantum mechanics the appropriate description of mixed states is by
density matrices.  For example, their compact notation makes them
useful for describing entanglement and decoherence properties of
multi-particle quantum systems.  In particular, two two-state density
matrices, also known as two qubit density matrices, are important for
their role in explaining quantum teleportation, dense coding,
computation theorems, and other issues pertinent to quantum
information theory.

Although the ideas behind extending classical computation and
communication theories into the quantum realm have been around for
some decades now, the first reference to calling any generic two-state
system a qubit comes from Schumacher \cite{Schumacher} in 1995.  By
calling a two-state system a qubit, he quantified the relationship
between classical and quantum information theory: a qubit can behave 
like a classical bit, but because of the quantum properties of
superposition and entanglement, it has a much larger information 
storage capacity.  It is this capacity to invoke quantum effects to 
increase information storage and processing, that gives qubits 
such a central role in quantum information theory.  

Now, a qubit is just a state in a two-dimensional Hilbert space
\cite{Preskill}.  If $\mathbb{H} \sim \mathbb{C}^2$ in the vector space, the
unit vectors
\begin{equation}
\ket{\psi}=a \ket{0}+b\ket{1}
\end{equation}
with a and b being complex numbers satisfying
\begin{equation}
\abs{a}^2+\abs{b}^2=1
\end{equation}
define, up to a phase, the pure quantum states.  In quantum
information theory, the orthonormal basis $\{\ket{0}, \ket{1}\}$ is
used to represent the bit states 0 (off) and 1 (on).  As pointed out
by Brown \cite{Brown} the physical representation of these two bit
states depends on the ``hardware'' being discussed; the basis states
may be polarization states of light, atomic or electronic spin states, 
or the ground and first excited state of a quantum dot.  

If the qubit represents a mixed state, which is quite often the case, one should use a
two-dimensional density matrix, which was introduced independently by
Landau and von Neumann in the 1920's (see for example the discussion in \cite{Davydov}),
for its representation.  The formalism of density matrices allows one to exploit simple matrix
algebra mechanisms to evaluate the
expectation value of any physical quantity of the system.
More recently, it has been pointed out by several people (see 
\cite{Preskill, MByrd4All, MByrd3Slater1, NielsonChuang},and references within) that the density matrix
representation of quantum states is also a very natural representation to use
with regards to quantum information calculations.  

Following this lead we therefore express one qubit as 
\begin{equation}
\rho=\frac{1}{2}(\Bid_2 + \boldsymbol{\sigma} \cdot \boldsymbol{n}),
\end{equation}
i.e.\ as a general 2 by 2 hermitian matrix with unit trace and the
positivity condition $Tr[\rho]\geq 0$ implying $\boldsymbol{n}\cdot
\boldsymbol{n} \leq 1$ or $\rho^2 \leq \rho$.  Therefore, these density matrices
are the disk $D^3$, whose boundary $\partial D^3 = S^2 = \mathbb{C}P^1$
represents the pure states ($\rho^2 = \rho$, or $\boldsymbol{n}\cdot
\boldsymbol{n} =1 $), and which can be thus characterized by the two
angles $0 \leq \theta \leq \pi$ (the latitude) and $0 \leq \phi \leq 2\pi$ (the longitude) of the
sphere $S^2$. 

Now, two-state density matrices live in a 3 by 3 hermitian-matrix
space, with $\mathbb{C}P^2 = SU(3)/U(2)$ as a subspace of pure
states.  Much is already known about these two- and three- state density matrices,
especially when one uses, for example, Euler angle parametrizations
(see \cite{MByrd3Slater1} for more information).  But what is not as well known is how the
density matrices of larger dimensional Hilbert spaces, and thus of 
multiple qubits, looks under such a parametrization.  This paper will make a great
deal of progress in 
remedying this situation by giving an explicit parametrization of the 
density matrix of two qubits that is not redundant in the 
representation of the corresponding four-dimensional Hilbert
space, and at the same time offers up the natural (Bures) volume measure 
on the set of all two qubit density matrices.  This will be achieved by
starting with a diagonal density matrix $\rho_{d}$, which represents our two 
qubit system in some particular basis, and then performing a unitary ($U^{-1} = U^\dagger$), unimodular
($\text{Det}[U] = 1$) transformation 
\begin{equation}
\rho_{d} \rightarrow U\rho_{d} U^{\dagger}
\end{equation} 
for some $U \in SU(4)$, thus describing $\rho$ in an 
arbitrary basis \cite{MByrd3Slater1, MByrd1, MByrdp1, MByrd2}.
We should 
point out that progress in this direction has also been made in 
a recent publication by Schlienz and Mahler \cite{Sclienz:95}.


\section{Euler Angle Parametrization for $SU(4)$}

We begin by giving the Euler angle parametrization for $SU(4)$.  
Define $U \in SU(4)$.  Using the Gell-Mann basis for the elements of the
algebra (found in Appendix \ref{app:crsu4}), the Euler angle
parametrization is then given by
\begin{eqnarray}
\label{su4eas}
U = e^{i\lambda_3 \alpha_1}e^{i\lambda_2 \alpha_2}e^{i\lambda_3 \alpha_3}e^{i\lambda_5 \alpha_4}e^{i\lambda_3 \alpha_5}e^{i\lambda_{10} \alpha_6}e^{i\lambda_3 \alpha_7}e^{i\lambda_2 \alpha_8}
e^{i\lambda_3 \alpha_{9}}e^{i\lambda_5 \alpha_{10}}e^{i\lambda_3 \alpha_{11}}e^{i\lambda_2 \alpha_{12}}e^{i\lambda_3 \alpha_{13}}e^{i\lambda_8 \alpha_{14}}e^{i\lambda_{15} \alpha_{15}}.
\end{eqnarray}
The derivation of this result is as follows.  We begin
by following the work of Biedenharn \cite{Biedenharn} and Hermann \cite{Hermann} in order to
generate a Cartan decomposition of $SU(4)$.  First, 
we look at the 4 by 4, hermitian, traceless, Gell-Mann matrices $\lambda_i$.
This set is linearly independent and is the lowest dimensional
faithful representation of the $SU(4)$ Lie algebra.  From these matrices we can
then calculate their commutation relations, and by observation of the
corresponding structure constants $f_{ijk}$ we can see the
relationship in the algebra that can help generate the Cartan
decomposition of $SU(4)$ (shown in detail in Appendix \ref{app:crsu4}).  

We now establish two subspaces of the $SU(4)$ group manifold hereafter
known as $K$ and $P$.  
From these subspaces, there corresponds two subsets of the Lie algebra of $SU(4)$, $L(K)$ and
$L(P)$, such that for $k_1, k_2 \in L(K)$ and $p_1, p_2 \in L(P)$,
\begin{align}
[k_1, k_2] &\in L(K), \nonumber\\
[p_1, p_2] &\in L(K), \nonumber\\
[k_1, p_2] &\in L(P). 
\end{align}
For $SU(4)$, $L(K) = \{ \lambda_{1},\ldots,\lambda_{8}, \lambda_{15} \}$  and $L(P) =
\{ \lambda_{9},\ldots,\lambda_{14} \}$.  Given that we can
decompose the $SU(4)$ algebra into a semi-direct sum \cite{Herstein}
\begin{equation}
L(SU(4))=L(K) \oplus L(P),
\end{equation}
we therefore have a decomposition of the group,
\begin{equation}
U=K\cdot P.
\end{equation}

From \cite{Sattinger} we know that $L(K)$ contains the generators of the
$SU(3)$ subalgebra of $SU(4)$, thus
$K$ will be the $U(3)$ subgroup obtained by
exponentiating the subalgebra,
$\{ \lambda_1,\ldots,\lambda_{8} \}$ combined with
$\lambda_{15}$ and thus can be written as (see \cite{MByrd1, MByrdp1} for details)
\begin{equation}
K=e^{i\lambda_3 \alpha}e^{i\lambda_2 \beta}e^{i\lambda_3 \gamma}
e^{i\lambda_5 \theta}e^{i\lambda_3 a}e^{i\lambda_2 b}
e^{i\lambda_3 c}e^{i\lambda_8 \chi}e^{i\lambda_{15} \phi}.
\end{equation}
Now, as for $P$, of the six elements in $L(P)$ we chose the
$\lambda_2$ analogue, $\lambda_{10}$, for $SU(4)$ and write any element of $P$ as
\begin{equation}
P=K'e^{i\lambda_{10}\psi}K''
\end{equation}
where $K'$ and $K''$ are copies of $K$. 

Unfortunately, at this point in our derivation, 
we have a $U$ with 28 elements, not the requisite 15
\begin{equation}
U=K K'e^{i\lambda_{10}\psi}K''.
\end{equation}
But, if we recall that $U$ is a product of operators in $SU(4)$, we
can ``remove the redundancies,'' i.e.\ the first $K'$ component as well as the three Cartan subalgebra elements of $SU(4)$ in the original $K$ component, 
to arrive at the following product \cite{MByrd1, MByrdp1}
\begin{eqnarray}
U = e^{i\lambda_3 \alpha_1}e^{i\lambda_2 \alpha_2}e^{i\lambda_3 \alpha_3}e^{i\lambda_5 \alpha_4}e^{i\lambda_3 \alpha_5}e^{i\lambda_2 \eta}e^{i\lambda_{10} \psi}e^{i\lambda_3 \alpha}e^{i\lambda_2 \beta}
e^{i\lambda_3 \gamma}e^{i\lambda_5 \theta}e^{i\lambda_3 a}e^{i\lambda_2 b}e^{i\lambda_3 c}e^{i\lambda_8 \chi}e^{i\lambda_{15} \phi}.
\end{eqnarray}
By insisting that our parametrization must truthfully reproduce known vector and tensor
transformations under $SU(4)$, we can remove the last ``redundancy,'' $e^{i\lambda_2 \eta}$, and, after rewriting the parameters, generate equation \eqref{su4eas}
\begin{eqnarray}
U = e^{i\lambda_3 \alpha_1}e^{i\lambda_2 \alpha_2}e^{i\lambda_3 \alpha_3}e^{i\lambda_5 \alpha_4}e^{i\lambda_3 \alpha_5}e^{i\lambda_{10} \alpha_6}e^{i\lambda_3 \alpha_7}e^{i\lambda_2 \alpha_8}
e^{i\lambda_3 \alpha_{9}}e^{i\lambda_5 \alpha_{10}}e^{i\lambda_3 \alpha_{11}}e^{i\lambda_2 \alpha_{12}}e^{i\lambda_3 \alpha_{13}}e^{i\lambda_8 \alpha_{14}}e^{i\lambda_{15} \alpha_{15}}.
\end{eqnarray}
For our purposes it is enough 
to note that this parametrization
is special unitary by construction and can 
be shown to cover the group by modifying the ranges that 
follow and substituting them into the 
whole group matrix, or into the parametrization of 
the characters \cite{Gibbons}.


\section{Derivation of the Haar Measure and Calculation of the Group Volume for $SU(4)$}

Taking the Euler angle parametrization given by equation \eqref{su4eas} we now
wish to develop the differential volume element, also known as the Haar
measure, for the group $SU(4)$.  We initially
proceed by extending the method used in \cite{MByrd1, MByrdp1} for the calculation
of the Haar measure for $SU(3)$; 
take a generic $U \in SU(4)$ and find the matrix
\begin{equation}
\label{initialform}
U^{-1} \cdot dU = U^{-1} \frac{\partial U}{\partial \alpha_{k}}\; d\alpha_{k}
\end{equation}
of left invariant one-forms, then wedge the 15 linearly independent
forms together.\footnote{Similarly one can wedge together 
the 15 right invariant one-forms which
also yields the Haar measure in question.  This
is due to the fact that a compact simply-connected real Lie group
has a bi-invariant measure, unique up to a constant factor.  Such a group
is usually referred to as `unimodular' \cite{Sattinger}.} 
But due to the 15 independent parameters needed for $SU(4)$, 
this method is unfortunately quite time consuming and thus 
prohibitive.  An easier way, initially given in \cite{Murnaghan}, is to calculate the
4 by 4 matrices, $\partial U / \partial \alpha_{k}$ (for $k=\{1\ldots
15\}$), and take the 
determinant of the coefficient matrix generated by their subsequent expansion in
terms of the Gell-Mann basis.

To begin, we take the transpose of equation \eqref{su4eas} to generate
\begin{eqnarray}
\label{su4conjeas}
u= U^T &=& e^{i\lambda_{15}^T \alpha_{15}}e^{i\lambda_{8}^T \alpha_{14}}e^{i\lambda_{3}^T \alpha_{13}}e^{i\lambda_{2}^T \alpha_{12}}e^{i\lambda_{3}^T \alpha_{11}}e^{i\lambda_{5}^T \alpha_{10}}e^{i\lambda_{3}^T \alpha_{9}}e^{i\lambda_{2}^T \alpha_{8}} \nonumber \\
&&\times e^{i\lambda_{3}^T \alpha_{7}}e^{i\lambda_{10}^T \alpha_{6}}e^{i\lambda_{3}^T \alpha_{5}}e^{i\lambda_{5}^T\alpha_{4}}e^{i\lambda_{3}^T \alpha_{3}}e^{i\lambda_{2}^T \alpha_{2}}e^{i\lambda_{3}^T \alpha_{1}}.
\end{eqnarray}
An observation of the components of our Lie algebra sub-set 
$(\lambda_2, \lambda_3, \lambda_5, \lambda_8, \lambda_{10}, \lambda_{15})$ 
shows that the transpose operation is equivalent to making the following substitutions
\begin{eqnarray}
\lambda_2^T \rightarrow -\lambda_2 &,& \lambda_3^T \rightarrow \lambda_3 ,\nonumber\\
\lambda_5^T \rightarrow -\lambda_5 &,& \lambda_8^T \rightarrow \lambda_8 ,\nonumber\\
\lambda_{10}^T \rightarrow -\lambda_{10} &,& \lambda_{15}^T \rightarrow \lambda_{15}
\end{eqnarray}
in equation \eqref{su4conjeas} generating
\begin{eqnarray}
u &=& e^{i\lambda_{15} \alpha_{15}}e^{i\lambda_{8} \alpha_{14}}e^{i\lambda_{3} \alpha_{13}}e^{-i\lambda_{2} \alpha_{12}}e^{i\lambda_{3}\alpha_{11}}e^{-i\lambda_{5} \alpha_{10}}e^{i\lambda_{3} \alpha_{9}}e^{-i\lambda_{2} \alpha_{8}} \nonumber \\
&&\times e^{i\lambda_{3} \alpha_{7}}e^{-i\lambda_{10}
\alpha_{6}}e^{i\lambda_{3} \alpha_{5}}e^{-i\lambda_{5} \alpha_{4}}e^{i\lambda_{3} \alpha_{3}}e^{-i\lambda_{2} \alpha_{2}}e^{i\lambda_{3} \alpha_{1}}.
\end{eqnarray}
Whichever form is used though, we then take the partial derivative of $u$ with respect to each of the
15 parameters.  
In general, the differentiation will have the form
\begin{eqnarray}
\label{partialgen}
\frac{\partial u}{\partial \alpha_{k}} &=& e^{i\lambda_{15}^T\alpha_{15}}e^{i\lambda_{8}^T \alpha_{14}}e^{i\lambda_{3}^T
\alpha_{13}}e^{i\lambda_{2}^T\alpha_{12}}\cdots
e^{i\lambda_{m}^T\alpha_{k+1}}{i\lambda_{n}^T}e^{i\lambda_{n}^T\alpha_{k}}e^{i\lambda_{p}^T\alpha_{k-1}}\cdots
e^{i\lambda_{3}^T\alpha_{1}} \nonumber \\
&=&
e^{i\lambda_{15}^T\alpha_{15}}e^{i\lambda_{8}^T\alpha_{14}}e^{i\lambda_{3}^T\alpha_{13}}e^{i\lambda_{2}^T\alpha_{12}}\cdots
e^{i\lambda_{m}^T\alpha_{k+1}}{i\lambda_{n}^T}
e^{-i\lambda_{m}^T\alpha_{k+1}}\cdots e^{-i\lambda_{2}^T\alpha_{12}}e^{-i\lambda_{3}^T\alpha_{13}}e^{-i\lambda_{8}^T \alpha_{14}}e^{-i\lambda_{15}^T\alpha_{15}}u
\end{eqnarray}
which, if we make the following definitions,
\begin{equation}
C(\alpha_k)\in i*\{ \lambda_{2}^T, \lambda_{3}^T, \lambda_{5}^T,
\lambda_{8}^T, \lambda_{10}^T, \lambda_{15}^T \},
\end{equation}
and
\begin{align}
E^{L}=&\;e^{C(\alpha_{15}) \alpha_{15}}\cdots e^{C(\alpha_{k+1}) \alpha_{k+1}},\nonumber\\
E^{-L}=&\;e^{-C(\alpha_{k+1}) \alpha_{k+1}}\cdots e^{-C(\alpha_{15}) \alpha_{15}}
\end{align}
can be expressed, in a ``shorthanded'' notation as
\begin{equation}
\frac{\partial u}{\partial \alpha_{k}}=E^{L}C(\alpha_k)E^{-L}u.
\end{equation}
By using these equations and the Baker-Campbell-Hausdorff relation,
\begin{equation}
 e^{X}Ye^{-X} = Y + [X,Y] + \frac{1}{2}[X,[X,Y]] + \ldots,
\end{equation}
we are able to consecutively solve equation \eqref{partialgen} for
$k=\{15,\ldots, 1\}$, giving us a set of 4 by 4 matrices which can be expanded in terms 
of the 15 \textit{transposed} elements of the $SU(4)$ Lie algebra with
expansion coefficients given by trigonometric functions of the group parameters $\alpha_i$:
\begin{equation}
\label{expansionmatrices}
M_{k}\equiv \frac{\partial u}{\partial \alpha_{k}}\; u^{-1} = E^{L}C(\alpha_k)E^{-L} = \sum_{15\ge j \ge 1} c_{kj}\lambda_j^T.
\end{equation}

At this stage, we should illustrate the connection between the
$M_{k}$'s and the 15 left invariant one-forms that we could have
used.  To begin we note that
\begin{gather}
du \cdot u^{-1} = d(U^T) \cdot (U^T)^{-1} = (dU)^T \cdot (U^{-1})^T = (U^{-1} \cdot dU)^T. \\
\intertext{Thus the following relationship between equation 
\eqref{initialform} and \eqref{expansionmatrices} holds:}
\biggl(\frac{\partial u}{\partial \alpha_{k}}\;d\alpha_{k}\biggr) u^{-1} =
\sum_{15\ge j \ge 1} c_{kj}\lambda_j^T d\alpha_{k} = (U^{-1}
\frac{\partial U}{\partial \alpha_{k}}\; d\alpha_{k})^T. \\
\intertext{Therefore we can conclude}
U^{-1} \frac{\partial U}{\partial \alpha_{k}}\; d\alpha_{k} =
\sum_{1\le j \le 15} c_{kj}\lambda_j d\alpha_{k}, 
\end{gather}
for $k=\{1,\ldots, 15\}$.  So even though we are calculating the \textit{right} invariant 
one-forms for $u$, we are really calculating the 
\textit{left} invariant one-forms for $U$.  The important thing to 
note is that the $c_{kj}$'s do not change.\footnote{The 
transpose operation on the Gell-Mann matrices only gives an overall sign difference to some of the matrices
\begin{gather}
\{\lambda_1^T, \lambda_2^T, \lambda_3^T, \lambda_4^T, \lambda_5^T, \lambda_6^T, \lambda_7^T, \lambda_8^T, \lambda_9^T, 
\lambda_{10}^T, \lambda_{11}^T, \lambda_{12}^T, \lambda_{13}^T, \lambda_{14}^T, \lambda_{15}^T\} \rightarrow \nonumber \\
\{\lambda_1, -\lambda_2, \lambda_3, \lambda_4, -\lambda_5, \lambda_6, -\lambda_7, \lambda_8, \lambda_9, -\lambda_{10},
\lambda_{11}, -\lambda_{12}, \lambda_{13}, -\lambda_{14}, \lambda_{15}\} \nonumber
\end{gather}
but these sign changes are augmented by the inversion in the sum over
k and therefore cancel out in the final construction of the
determinant that we need.  The transpose operation on equation
\eqref{su4eas} is done only to simplify the initial evaluation of the
expansion coefficients $c_{kj}$. }

Now, the expansion coefficients $c_{kj}$ are the elements of the determinant 
in question.  They are found in the following manner
\begin{equation}
c_{kj} = \frac{-i}{2}\;Tr[\lambda_j^T \cdot M_{k}]
\end{equation}
where the trace is done over all 15 \textit{transposed} Gell-Mann matrices \cite{Greiner}.
The index $k$ corresponds to the specific $\alpha$ parameter and the $j$ corresponds to the specific element of the algebra.  Both the $k$ and $j$ indices run from 15 to 1.
The determinant of this 15 by 15 matrix yields the differential volume
element, also known as the Haar measure for the group, $dV_{SU(4)}$
that, when integrated over the correct values for the ranges of the
parameters and multiplied by a derivable normalization constant,
yields the volume for the group.

The full 15 by 15 determinant $\text{Det}[c_{kj}]$, $k,j \in \{15,\ldots,1\}$, can be done, or one can notice that the
determinant can be written as
\begin{equation}
C_{SU(4)}= \begin{Vmatrix}
  c_{15,14} & c_{15,13} &\dots& c_{15,1} & c_{15,15}\\
  c_{14,14} & c_{14,13} &\dots& c_{14,1} & c_{14,15}\\
  \dots & \dots & \dots & \dots & \dots \\
  c_{1,14} & c_{1,13} &\dots& c_{1,1} & c_{1,15}
\end{Vmatrix}
\end{equation}
which differs only by an overall sign from $\text{Det}[c_{kj}]$ above, but
which also yields a quasi-block form that generates
\begin{equation}
C_{SU(4)} =\begin{Vmatrix}
  O & D \\
  A & B 
\end{Vmatrix},
\end{equation}
where $D$ corresponds to the 9 by 9 matrix whose determinant is
equivalent to 
$dV_{SU(3)}\cdot d\alpha_{15}$ \cite{MByrd1}, $B$ is a complicated 6 by 9
matrix, and $O$ is a 9 by 6 matrix
whose elements are all zero.  

Now the interchange of
two columns of a $N$ by $N$ matrix yields a change in sign
of the corresponding determinant, but by moving six columns at once the
sign of the determinant does not change and one may
therefore generate a new matrix
\begin{equation}
C'_{SU(4)}= \begin{Vmatrix}
  c_{15,8} & c_{15,7} &\dots& c_{15,1} & c_{15,15} & c_{15,14} & c_{15,13} &\dots&
  c_{15,9}\\
  c_{14,8} & c_{14,7} &\dots& c_{14,1} & c_{14,15} & c_{14,14} & c_{14,13} &\dots&
  c_{14,9}\\
  \dots & \dots & \dots & \dots & \dots & \dots & \dots & \dots & \dots\\
  c_{1,8} & c_{1,7} &\dots& c_{1,1} & c_{1,15} & c_{1,14} & c_{1,13} &\dots& c_{1,9}
\end{Vmatrix},
\end{equation} 
which is now block diagonal
\begin{equation}
C'_{SU(4)} =\begin{Vmatrix}
  D & O \\
  B & A 
\end{Vmatrix},
\end{equation}
and which yields the same determinant as $C_{SU(4)}$.  Thus, with this
new form, the full determinant is just equal to the determinant of the
diagonal blocks, one of which is already known.\footnote{The general determinant 
formula for this type of block matrix is given, without proof, by Robert Tucci in \eprint{quant-ph$\backslash$0103040}.}  
So only the determinant of the 6 by 6
sub-matrix $A$ which is equal to
\begin{equation}
 A = \begin{Vmatrix}
  c_{6,14} & c_{6,13} &\dots& c_{6,10} & c_{6,9}\\
  c_{5,14} & c_{5,13} &\dots& c_{5,10} & c_{5,9}\\
  \dots & \dots & \dots & \dots & \dots \\
  c_{1,14} & c_{1,13} &\dots& c_{1,10} & c_{1,9}
\end{Vmatrix}
\end{equation}
is needed.  Therefore the differential volume element for $SU(4)$ is nothing more than
\begin{eqnarray}
dV_{SU(4)} &=&\text{Det}[c_{kj}]\nonumber\\ 
&=& -\text{Det}[A] * \text{Det}[D]d\alpha_{15}\ldots d\alpha_{1}\nonumber\\
&=& -\text{Det}[A] * dV_{SU(3)}d\alpha_{15}d\alpha_{6}\ldots d\alpha_{1}
\end{eqnarray}
which when calculated yields the Haar measure
\begin{eqnarray}
\label{dvsu4}
dV_{SU(4)} &=& \cos(\alpha_{4})^3\cos(\alpha_{6})\cos(\alpha_{10})\sin(2\alpha_{2})\sin(\alpha_{4})
\sin(\alpha_{6})^5\sin(2\alpha_{8})\sin(\alpha_{10})^3\sin(2\alpha_{12})d\alpha_{15}\ldots d\alpha_{1}.
\end{eqnarray}
This is determined up to normalization (explained in detail in
Appendix \ref{app:Haar}).  Integration over the 15 parameter space gives the 
group volume 
\begin{eqnarray}
\idotsint\limits_V dV_{SU(4)} &=& (192)\idotsint\limits_{V^\prime} dV_{SU(4)}\nonumber\\
V_{SU(4)} &=& \frac{\sqrt{2}\pi^9}{3},
\end{eqnarray}
which is in agreement with the volume obtained by
Marinov \cite{Marinov2}.


\section{Two Qubit Density Matrix Parametrization}

Using this Euler angle parametrization, any two qubit density matrix
can now be represented by following the convention derived in Boya 
et.\ al. \cite{MByrd4All}.  As stated in Boya et.\ al., any $N$-dimensional pure state can be
written as a diagonal matrix with one element equal to 1 and the rest
zero.  Different classes of pure states have a different
ordering of the zero and non-zero diagonal elements.  Therefore, if one
wants to write down a mixture of these different pure states, one must
take the following convex sum
\begin{equation}
\rho_d = \sum_i a^i \rho_i,
\end{equation}
where $\rho_d$ is now the mixed state, $\rho_i$ ($i$ running from 1 to $N$) are the
pure state matrices satisfying $Tr[\rho_i \rho_j]=2\delta_{ij}$, and
the $a^i$ are constants that satisfy $\sum_i a^i =1$ and $0 \leq a^i
\leq 1$ \cite{MByrd4All}.  Now the $a^i$ are just the eigenvalues of the
density matrix $\rho_d$ and can thus be parameterized by the squared
components within the ($N-1$) sphere, $S^{N-1}$.  If we now want the most general mixed
state density matrix in some arbitrary basis, one only has to perform a unitary, unimodular
transformation upon $\rho_d$; a transformation that will be an
element of $SU(N)$.  So for our two qubit density matrix $\rho$ we write
\begin{equation}
\label{rhodm}
\rho = U \rho_d U^\dagger,
\end{equation}
where $\rho_d$ is the diagonalized density matrix which corresponds 
to the eigenvalues of the 3-sphere, $S^3$ \cite{MByrd4All, MByrd3Slater1, MByrd2}
\begin{equation}
\label{rhod}
\rho_d = \left( \begin{matrix}
\sin^2(\theta_1)\sin^2(\theta_2)\sin^2(\theta_3) & 0 & 0 & 0\\
0 & \cos^2(\theta_1)\sin^2(\theta_2)\sin^2(\theta_3) & 0 & 0\\
0 & 0 & \cos^2(\theta_2)\sin^2(\theta_3) & 0 \\
0 & 0 & 0 & \cos^2(\theta_3) 
\end{matrix} \right)
\end{equation}
 and $U$, now an element of $SU(4)$, is from equation \eqref{su4eas}.

It is instructive to rewrite equation \eqref{rhod} as
the exponentiated product of generators of the Cartan subalgebra that we are
using in our parametrization of $SU(4)$; $e^{\lambda_3*a}, e^{\lambda_8*b},\text{ and }
e^{\lambda_{15}*c}$.  
Unfortunately, indeterminacies with the logarithm of the elements of $\rho_d$ does not
allow for such a rewrite so $\rho_d$ will be expressed in terms of the following sum
\begin{equation}
\sum_{1 \leq j \leq 15}w_j\lambda_j+w_0\Bid_{4}.
\end{equation}
We begin by redefining $\rho_d$ in the
following way
\begin{equation}
\label{rhod2}
\rho_d = \left(\begin{matrix}
         w^2x^2y^2 & 0 & 0 & 0 \\
         0 & (1-w^2)x^2y^2 & 0 & 0  \\
         0 & 0 & (1-x^2)y^2 & 0 \\
         0 & 0 & 0 & 1-y^2
\end{matrix}\right),
\end{equation}
where $w^2 = \sin^2(\theta_1)$, $x^2 = \sin^2(\theta_2)$, and
$y^2=\sin^2(\theta_3)$.  Now we calculate the decomposition of equation \eqref{rhod2}
in terms of the elements of the full Lie algebra.  This is accomplished by
taking the trace of $\frac{1}{2}\rho_d \cdot \lambda_j$ over all 15 Gell-Mann matrices.
Evaluation of these 15 trace operations yields the following
decomposition of equation \eqref{rhod2}
\begin{eqnarray}
\label{rhofactor}
\rho_d &=& \frac{1}{4}\Bid_{4} +
\frac{1}{2}(-1+2w^2)x^2y^2*\lambda_3 
+ \frac{1}{2\sqrt{3}}(-2+3x^2)y^2*\lambda_8 +
\frac{1}{2\sqrt{6}}(-3+4y^2)*\lambda_{15},
\end{eqnarray}
where the one-quarter $\Bid_{4}$ keeps the trace of $\rho_d$ in this form still unity.

With equations \eqref{rhodm} and \eqref{rhofactor} we can write
down $\rho$ completely in terms of the Lie algebra sub-set of
the parametrization.  First, $U^\dagger$, the
transpose of the conjugate of equation \eqref{su4eas}, is
expressed as
\begin{eqnarray}
U^\dagger &=& e^{-i\lambda_{15} \alpha_{15}}e^{-i\lambda_{8} \alpha_{14}}e^{-i\lambda_{3} \alpha_{13}}e^{-i\lambda_{2} \alpha_{12}}e^{-i\lambda_{3} \alpha_{11}}e^{-i\lambda_{5} \alpha_{10}}e^{-i\lambda_{3} \alpha_{9}}e^{-i\lambda_{2} \alpha_{8}} \nonumber \\
&&\times e^{-i\lambda_{3} \alpha_{7}}e^{-i\lambda_{10} \alpha_{6}}e^{-i\lambda_{3} \alpha_{5}}e^{-i\lambda_{5} \alpha_{4}}e^{-i\lambda_{3} \alpha_{3}}e^{-i\lambda_{2} \alpha_{2}}e^{-i\lambda_{3} \alpha_{1}}.
\end{eqnarray}
Thus equation \eqref{rhodm} is equal to
\begin{eqnarray}
\rho &=& e^{i\lambda_3 \alpha_1}e^{i\lambda_2 \alpha_2}e^{i\lambda_3 \alpha_3}e^{i\lambda_5 \alpha_4}e^{i\lambda_3 \alpha_5}e^{i\lambda_{10} \alpha_6}e^{i\lambda_3 \alpha_7}e^{i\lambda_2 \alpha_8}
\nonumber \\
& & \times e^{i\lambda_3 \alpha_{9}}e^{i\lambda_5 \alpha_{10}}e^{i\lambda_3 \alpha_{11}}e^{i\lambda_2
\alpha_{12}}e^{i\lambda_3 \alpha_{13}}e^{i\lambda_8 \alpha_{14}}e^{i\lambda_{15} \alpha_{15}}\nonumber\\
& & \times (\frac{1}{4}\Bid_{4}+\frac{1}{2}(-1+2w^2)x^2y^2*\lambda_3 +
\frac{1}{2\sqrt{3}}(-2+3x^2)y^2*\lambda_8+
\frac{1}{2\sqrt{6}}(-3+4y^2)*\lambda_{15})\nonumber\\
& & \times e^{-i\lambda_{15} \alpha_{15}}e^{-i\lambda_{8} \alpha_{14}}e^{-i\lambda_{3} \alpha_{13}}e^{-i\lambda_{2} \alpha_{12}}e^{-i\lambda_{3} \alpha_{11}}e^{-i\lambda_{5} \alpha_{10}}e^{-i\lambda_{3} \alpha_{9}}e^{-i\lambda_{2} \alpha_{8}}
\nonumber \\
& &\times e^{-i\lambda_{3} \alpha_{7}}e^{-i\lambda_{10} \alpha_{6}}e^{-i\lambda_{3} \alpha_{5}}e^{-i\lambda_{5}
\alpha_{4}}e^{-i\lambda_{3} \alpha_{3}}e^{-i\lambda_{2} \alpha_{2}}e^{-i\lambda_{3} \alpha_{1}},
\end{eqnarray}
which, because $\Bid_{4}$, $\lambda_3$, $\lambda_8$, and $\lambda_{15}$ all
commute with each other, has the following simplification
\begin{eqnarray}
\rho &=& \cdots e^{i\lambda_3 \alpha_{13}}e^{i\lambda_8
  \alpha_{14}}e^{i\lambda_{15} \alpha_{15}} \nonumber\\
& &\times(\frac{1}{4}\Bid_{4}+\frac{1}{2}(-1+2w^2)x^2y^2*\lambda_3
+\frac{1}{2\sqrt{3}}(-2+3x^2)y^2*\lambda_8+
\frac{1}{2\sqrt{6}}(-3+4y^2)*\lambda_{15}) \nonumber \\
& &\times e^{-i\lambda_{15} \alpha_{15}}e^{-i\lambda_{8}
  \alpha_{14}}e^{-i\lambda_{3} \alpha_{13}}\cdots\\
& = &\cdots (\frac{1}{4}\Bid_{4}+\frac{1}{2}(-1+2w^2)x^2y^2*\lambda_3
+\frac{1}{2\sqrt{3}}(-2+3x^2)y^2*\lambda_8+
\frac{1}{2\sqrt{6}}(-3+4y^2)*\lambda_{15})\cdots\nonumber
\end{eqnarray}
Therefore, all density matrices in $SU(4)$ have the following form\footnote{One 
should be able to write only 12 matrices in the parametrization of $U$ since the 
little group of $\rho_d$ is generated by $\lambda_3, \lambda_8, \text{ and }\lambda_{15}$.}
\begin{eqnarray}
\label{uberrho}
\rho &=& e^{i\lambda_3 \alpha_1}e^{i\lambda_2 \alpha_2}e^{i\lambda_3 \alpha_3}e^{i\lambda_5 \alpha_4}e^{i\lambda_3 \alpha_5}e^{i\lambda_{10} \alpha_6}e^{i\lambda_3 \alpha_7}e^{i\lambda_2 \alpha_8}
e^{i\lambda_3 \alpha_{9}}e^{i\lambda_5 \alpha_{10}}e^{i\lambda_3 \alpha_{11}}e^{i\lambda_2 \alpha_{12}} \nonumber \\
& &\times (\frac{1}{4}\Bid_{4}+\frac{1}{2}(-1+2w^2)x^2y^2*\lambda_3 +
\frac{1}{2\sqrt{3}}(-2+3x^2)y^2*\lambda_8+\frac{1}{2\sqrt{6}}(-3+4y^2)*\lambda_{15})\nonumber\\
& &\times e^{-i\lambda_{2} \alpha_{12}}e^{-i\lambda_{3} \alpha_{11}}e^{-i\lambda_{5} \alpha_{10}}e^{-i\lambda_{3} \alpha_{9}}
e^{-i\lambda_{2} \alpha_{8}}e^{-i\lambda_{3} \alpha_{7}}e^{-i\lambda_{10} \alpha_{6}}e^{-i\lambda_{3} \alpha_{5}}e^{-i\lambda_{5}
\alpha_{4}}e^{-i\lambda_{3} \alpha_{3}}e^{-i\lambda_{2} \alpha_{2}}e^{-i\lambda_{3} \alpha_{1}},
\end{eqnarray}
where
\begin{eqnarray}
\label{factorlist}
w^2 &=& \sin^2(\theta_1),\nonumber\\
x^2 &=& \sin^2(\theta_2),\nonumber\\
y^2 &=& \sin^2(\theta_3),
\end{eqnarray}
with the ranges for the 12 $\alpha$ parameters and the 
three $\theta$ parameters given by
\begin{gather}
0 \le \alpha_1,\alpha_3,\alpha_5,\alpha_7,\alpha_9,\alpha_{11} \le
\pi, \nonumber \\
0 \le \alpha_2,\alpha_4,\alpha_6,\alpha_8,\alpha_{10},\alpha_{12} \le
\frac{\pi}{2}, \nonumber \\
\begin{alignat}{3}
\frac{\pi}{4} \le \theta_1 \le \frac{\pi}{2}, \quad &
\cos^{-1}(\frac{1}{\sqrt{3}}) \le \theta_2 \le \frac{\pi}{2}, \quad &
\frac{\pi}{3} \le \theta_3 \le \frac{\pi}{2}.
\end{alignat}
\end{gather}
where the $\alpha$'s are determined from calculating the volume 
for the group (explained in Appendix \ref{app:Haar}) and the 
$\theta$'s from generalizing the work contained in \cite{MByrd3Slater1}.

In this manner all two-particle bipartite systems can be described by
a $\rho$ that is parameterized 
using 12 Euler angles, and three spatial rotations, and which by
\cite{HornJohnson}, majorizes all other density matrices of
$SU(4)$.\footnote{The eigenvalues of the given $\rho$ always satisfy $\nu_1 \geq
(\nu_2, \nu_3, \nu_4)$ with additional ordering between the $\nu_2,
\nu_3$ and $\nu_4$ eigenvalues.  Therefore one can always find an ordering of
the $\nu_i$ that satisfies the majorization condition.} 
Exploitations of this property, related to Birkhoff's theorem
concerning doubly stochastic matrices and convex sets
\cite{HornJohnson}, allows us to
use this parametrization to find the subset of ranges that generate
entangled density matrices and thus parameterize the convex polygon
that describes the set of entangled two qubit systems in terms of
Euler angles and spatial rotations.  In order to do this, we need to look at the partial transpose
of equation \eqref{uberrho}.\footnote{It is worth noting that Englert and Metwally 
\cite{Englert1, Englert2} have shown that for certain 
purposes, 9 parameters extracted from the density matrix are enough to 
describe certain important characteristics of the local and nonlocal 
properties of the density matrix.  In such cases, one should use the 
density matrix representation discussed in section \ref{sec:su4conc} which
more clearly expresses their ideas.}


\section{Reformulated Partial Transpose Condition}

To begin, one could say that a particular operation provides
\textit{some} entanglement if the 
following condition holds.  Let $\rho$ be a density matrix composed of two 
pure separable qubit states.  Then the following matrix will represent 
the two qubit subsystems $A$ and $B$,
\begin{equation}
\rho = \rho_A \otimes \rho_B.
\end{equation}
Let $U \in SU(4)$ be a matrix transformation 
on two qubits.  Therefore, if
\begin{equation}
\rho^\prime = U\rho U^\dagger
\end{equation}
is an entangled state, then the operation is capable of producing 
entanglement \cite{Zanardi00:1, Zanardi01:1}.  One way in which we can tell that the matrix 
$\rho^\prime$ is entangled, is to take the partial transpose of the 
matrix and see if it is positive (this is the Peres-Horodecki 
criterion \cite{Peres, Horodeckietal}).
In other words we wish to see if 
\begin{equation}
(\rho^\prime)^{T_A} \leq 0, \quad \text{or} \quad 
(\rho^\prime)^{T_B} \leq 0.
\end{equation}
These relations imply each of the partial transposes, $T_{A} \text{
  and } T_{B}$, leaves $\rho$ non-negative.
If either of these conditions are met, then there is entanglement.

As an example of this we look at the situation where $\rho_d=\rho$ and $U$ is given by equation
\eqref{su4eas}.  By taking the partial transpose of $\rho'$ and finding the 
subset of the given ranges of $\rho_d$ and $U$ such that
$\rho'$ satisfies the above conditions for entanglement 
we will be able to derive the set of all matrices which
describes the entanglement of two qubits.  To do this, we 
look at the eigenvalues of the partial transpose of $\rho'$.

Using the Euler angle parametrization previously given, a numerical
calculation of the eigenvalues of the partial transpose of $\rho'$
has been attempted.  Under the standard Peres-Horodecki criterion, if any of the
eigenvalues of the partial transpose of $\rho'$ are negative, then we
have an entangled $\rho'$ otherwise the state $\rho'$ is
separable.  As we have mentioned, we would like to derive a subset of the ranges of
the Euler angle parameters involved that would yield such a situation,
thus dividing the 15 parameter space into entangled and separable
subsets.  Unfortunately, due to the complicated nature of the 
parametrization, both numerical and symbolic calculations of the eigenvalues of the partial
transpose of $\rho'$ have become computationally intractable using
standard mathematical software.  Therefore, only a limited 
number of searches over the 15 parameter space
of those parameter values that satisfy the Peres-Horodecki
criterion have been attempted.  These initial calculations, though,
have shown that all possible combinations of the minimum and maximum
values for the 12 $\alpha$ and three $\theta$ parameters do not yield
entangled density matrices.  Numerical work has also shown that with
this parametrization, one, and only one, eigenvalue will be negative
when the values of the parameters give entangled density matrices.
This is a verification of Sanpera et.\ al.\ and Verstraete et.\ al.\ 
who have shown that the
partial transpose of an entangled two-qubit state is always of full
rank and has at most one negative eigenvalue \cite{Sanpera, Verstraete}.
This result is important, for it allows us to move away from using the standard
Peres-Horodecki criterion and substitute it with an expression that
only depends on the sign of a determinant.

To begin, the eigenvalue equation for a 4 by 4 matrix is of the form
\begin{equation}
(\lambda - \mu_1)(\lambda - \mu_2)(\lambda - \mu_3)(\lambda - \mu_4),
\end{equation}
which generates
\begin{equation}
\label{eigen}
\lambda^4+a\lambda^3+b\lambda^2+c\lambda+d=0,
\end{equation}
where
\begin{eqnarray}
\label{abcd}
a &=& -(\mu_1 + \mu_2 + \mu_3 + \mu_4),\nonumber\\
b &=& \mu_1\mu_2 + \mu_1\mu_3 + \mu_1\mu_4 + \mu_2\mu_3 + \mu_2\mu_4
+ \mu_3\mu_4,\nonumber\\
c &=& -(\mu_1\mu_2(\mu_3 + \mu_4) + (\mu_1 +
\mu_2)\mu_3\mu_4),\nonumber\\
d &=& \mu_1\mu_2\mu_3\mu_4.
\end{eqnarray}
Now, since the $\mu_i$ are eigenvalues, there sum must equal 1.  Thus,
coefficient \textit{a} in equation \eqref{eigen} is -1.  Therefore the
characteristic equation we
must solve is given by
\begin{equation}
\label{basic4}
\lambda^4-\lambda^3+b\lambda^2+c\lambda+d=0
\end{equation}
which can be simplified by making the substitution $\tau=\lambda-1/4$
which yields
\begin{eqnarray}
\label{quartic}
\tau^4+p\tau^2+q\tau+r&=&0,\nonumber\\
p&=&b-\frac{3}{8},\nonumber\\
q&=&\frac{b}{2}-c-\frac{1}{8},\nonumber\\
r&=&\frac{b}{16}-\frac{c}{4}+d-\frac{3}{256}.
\end{eqnarray}
The behavior of the solutions of this equation depends on the cubic resolvent
\begin{equation}
\label{crbasic}
\gamma^3 + 2p\gamma^2 + (p^2-4r)\gamma - q^2 = 0,
\end{equation}
which has $\gamma_{1}\gamma_{2}\gamma_{3}=q^2$ \cite{HoM}.  Recalling
that the solution of a cubic equation can be obtained by using
Cardano's formula \cite{HoM} we can immediately write down the roots
for equation \eqref{crbasic}
\begin{align}
\gamma_{1}=&\frac{-2p}{3}-\frac{2^{\frac{1}{3}}(-p^2-12r)}{3(2p^3+27q^2-72pr+\sqrt{4(-p^2-12r)^3+(2p^3+27q^2-72pr)^2})^{\frac{1}{3}}}+\nonumber\\
&\frac{(2p^3+27q^2-72pr+\sqrt{4(-p^2-12r)^3+(2p^3+27q^2-72pr)^2})^{\frac{1}{3}}}{32^{\frac{1}{3}}},\nonumber\\
\gamma_{2}=&\frac{-2p}{3}+\frac{(1+i\sqrt{3})(-p^2-12r)}{32^{\frac{2}{3}}(2p^3+27q^2-72pr+\sqrt{4(-p^2-12r)^3+(2p^3+27q^2-72pr)^2})^{\frac{1}{3}}}-\nonumber\\
&\frac{(1-i\sqrt{3})(2p^3+27q^2-72pr+\sqrt{4(-p^2-12r)^3+(2p^3+27q^2-72pr)^2})^{\frac{1}{3}}}{62^{\frac{1}{3}}},\nonumber\\
\gamma_{3}=&\frac{-2p}{3}+\frac{(1-i\sqrt{3})(-p^2-12r)}{32^{\frac{2}{3}}(2p^3+27q^2-72pr+\sqrt{4(-p^2-12r)^3+(2p^3+27q^2-72pr)^2})^{\frac{1}{3}}}-\nonumber\\
&\frac{(1+i\sqrt{3})(2p^3+27q^2-72pr+\sqrt{4(-p^2-12r)^3+(2p^3+27q^2-72pr)^2})^{\frac{1}{3}}}{62^{\frac{1}{3}}}.
\end{align}
In terms of our original parameters $b$, $c$, and $d$, we have for
equation \eqref{crbasic} 
\begin{equation}
\gamma^3+2(\frac{-3}{8}-b)\gamma^2+(\frac{3}{16}-b+b^2+c-4d)\gamma-\frac{1}{64}(1-4b+8c)^2,
\end{equation}
and therefore for its roots
\begin{align}
\gamma_{1}=&\; -12( -1 + 8b )( -3 + 16b( 1 + b )  - 16c + 64d )
+[ -54( 1 - 8b )^2( 1 - 4b + 8c )^2 + 6\sqrt{3} \nonumber \\
&\times \smash[t]{\sqrt{27( 1 - 8b )^4 ( 1 - 4b + 8c )^4 + 
16( -1 + 8b )^3( -3 + 16b( 1 + b )  - 16c + 64d )^3}}] 
^\frac{2}{3} \nonumber \\
&\times (12( -1 + 8b ) 
[ -54( 1 - 8b )^2( 1 - 4b + 8c )^2 + 6\sqrt{3} \nonumber\\
&\times \smash[t]{\sqrt{27( 1 - 8b )^4( 1 - 4b + 8c )^4 + 
16( -1 + 8b )^3( -3 + 16b( 1 + b )  - 16c + 64d )^3}}] 
^{\frac{1}{3}})^{-1}, \nonumber \\
\gamma_{2}=&\; 6( 1 + i\sqrt{3} ) ( -1 + 8b )( -3 + 16b( 1 + b )  - 16c + 64d )
+ ( -3 )^\frac{2}{3}
[ -18( 1 - 8b )^2 ( 1 - 4b + 8c )^2 + 2\sqrt{3} \nonumber\\
&\times \sqrt{27( 1 - 8b )^4( 1 - 4b + 8c )^4 + 
16( -1 + 8b )^3( -3 + 16b( 1 + b )  - 16c + 64d )^3}]
^\frac{2}{3} \nonumber\\
&\times (12( -1 + 8b ) 
[ -54( 1 - 8b )^2( 1 - 4b + 8c )^2 + 6\sqrt{3} \nonumber\\
&\times \sqrt{27( 1 - 8b )^4 ( 1 - 4b + 8c )^4 + 
16( -1 + 8b )^3( -3 + 16b( 1 + b )  - 16c + 64d )^3}] 
^\frac{1}{3})^{-1}, \nonumber \\
\gamma_{3}=&\; 12( 1 - i {\sqrt{3}} ) ( -1 + 8b ) 
( -3 + 16b( 1 + b )  - 16c + 64d )
- 3^{\frac{1}{6}}( -3i  + {\sqrt{3}} ) 
{[} -18{( 1 - 8b ) }^2{( 1 - 4b + 8c ) }^2 + 2{\sqrt{3}}\nonumber\\
&\times {\sqrt{27{( 1 - 8b ) }^4{( 1 - 4b + 8c ) }^4 + 
16{( -1 + 8b ) }^3{( -3 + 16b( 1 + b )  - 16c + 64d ) }^3}}{]} 
^{\frac{2}{3}}\nonumber\\
&\times {(}24( -1 + 8b ) {[} -54{( 1 - 8b ) }^2
{( 1 - 4b + 8c ) }^2 + 6{\sqrt{3}}\nonumber\\
&\times{\sqrt{27{( 1 - 8b ) }^4{( 1 - 4b + 8c ) }^4 + 
16{( -1 + 8b ) }^3{( -3 + 16b( 1 + b )  - 16c + 64d ) }^3}} {]}
^{\frac{1}{3}} {)}^{-1}.
\label{eq:biggamma}
\end{align}

Now, if all three $\gamma$ solutions are
real and positive, the quartic equation \eqref{quartic} has the following solutions
\begin{eqnarray}
\label{quarticsolutions}
\tau_{1}=\frac{\sqrt{\gamma_{1}}+\sqrt{\gamma_{2}}+\sqrt{\gamma_{3}}}{2},&\quad&
\tau_{2}=\frac{\sqrt{\gamma_{1}}-\sqrt{\gamma_{2}}-\sqrt{\gamma_{3}}}{2},\nonumber\\
\tau_{3}=\frac{-\sqrt{\gamma_{1}}+\sqrt{\gamma_{2}}-\sqrt{\gamma_{3}}}{2},&\quad&
\tau_{4}=\frac{-\sqrt{\gamma_{1}}-\sqrt{\gamma_{2}}+\sqrt{\gamma_{3}}}{2}.
\end{eqnarray}
Substitution of the $\gamma$ values given in equation (\ref{eq:biggamma})
into equation \eqref{quarticsolutions} creates 
the four eigenvalue equations that the standard Peres-Horodecki
criterion would force us to evaluate.
These are quite difficult and time consuming, 
especially when $b$, $c$, and $d$ are written
in terms of the twelve $\alpha$ and three $\theta$ parameters, and can
become computationally intractable even for modern mathematical software.
But, from the previous discussion, 
it is obvious that with only one
eigenvalue that changes sign, the only parameter that needs to be
analyzed is $d$.  Therefore, instead of looking at 
solutions of \eqref{quarticsolutions} one may instead look at when
$d$ from equation \eqref{abcd} changes sign.\footnote{A. Wang
has proposed a general solution the eigenvalue problem for the partial
transpose of two qubits (see for example \cite{Wang}
eq.\ (22)) in which he states that only one equation need be evaluated
to determine entanglement.
Unfortunately, in order to evaluate that one equation (\cite{Wang}
eq.\ (22)), six other equations must first be evaluated
(\cite{Wang} eqs.\ (23$\sim$28)).  In terms of the
15 parameters needed to represent the Hilbert space of a two qubit
density matrix, it is far easier to evaluate the zeroth order $\lambda$
coefficient $d$ given in equation \eqref{basic4} than it is to
evaluate seven total equations.  Even if one where to substitute 
and simplify, achieving one equation, its representation in terms of the 15
parameters needed to accurate describe the most general density matrix
would still be more complicated to numerically and symbolic evaluate
than the $d$ parameter.} 

Now, the $d$ parameter is the zeroth order $\lambda$ coefficient from
the following equation
\begin{equation}
\label{detc}
\text{Det}(\rho^{pt}-\Bid_{4}*\lambda) 
\end{equation} 
where $\rho^{pt}$ is the
partial transpose of equation \eqref{uberrho}.  This is just the
standard characteristic equation that yields the fourth-order
polynomial from which the eigenvalues of the partial transpose of
equation \eqref{uberrho} are to be evaluated, and which equations
\eqref{eigen} and \eqref{abcd} are generated from.  Computationally, 
from the standpoint of our parametrization, it is easier to take this
determinant than it is to explicitly solve for the roots of a
fourth-order polynomial (as we have given above).  The solution of
equation \eqref{detc} yields an expression for $d$ in terms of the 12
$\alpha$ and three $\theta$ parameters that
can be numerically evaluated by standard mathematical software
packages with much greater efficiency than the full Peres-Horodecki
criterion.\footnote{This greater efficiency is based on the observation 
that the kernel of the mathematical software package
\textsc{Mathematica} ver.\ 4.0 rel.\ 3, running on an optimized 1.5 GHz 
Pentium 4 Linux box with 1 gigabyte of 333MHz DDR, was unable to
express equation \eqref{quarticsolutions} in terms of the 12 $\alpha$ 
and three $\theta$ parameters in a format suitable for 
encoding into a C++ program.  On the other hand, it was quite easy to 
obtain all the coefficients of equation \eqref{basic4} in terms 
of our Euler parameters, simplify them, and encoded them into a C++ 
program for numerical evaluation.  Also, since only one eigenvalue 
ever goes negative, we have reduced the number of equations to solve 
from 4 to 1 which is a definite improvement in calculatory efficiency.}


\section{Conclusions}
\label{sec:su4conc}
The aim of this paper has been to show an explicit Euler angle 
parametrization for the Hilbert space of all two qubit density matrices.  
As we have stated, such a parametrization should 
be very useful for many calculations, especially numerical, concerning
entanglement.
This parametrization also allows for an in-depth 
analysis of the convex sets, sub-sets, and overall set boundaries of
separable and entangled two qubit systems without having to make any
initial restrictions as to the type of 
parametrization and density matrix in question.  
We have also been able to use this parametrization as an independent verification to Marinov's $SU(4)$ volume
calculation.
The role of the parametrization in simplifying the
Peres-Horodecki criteria 
for two qubit systems has also be indicated.  

Although one may
generate or use other parametrizations of $SU(4)$ and 
two qubit density matrices (see for example \cite{Marinov, Glasser, Zyczkowski1, VK}) our
parametrization does have the
advantage of not naively overcounting the group, as well as generating
an the easily integrable Haar measure and having a form
suited for generalization.  
Such a parametrization should also assist in 
providing a Bures distance for the space of two-qubits.
Also, although previous work has been
done on evaluating the eigenvalues of the partial transpose of the
two qubit density matrix (for example the work done by Wang in
\cite{Wang}), our representation allows the 
user to effect both a reduction in the number of equations to be
analyzed for entanglement onset from 4 to 1 while still retaining the
ability to analyze the little group and orbit space of the density
matrix as well (see for example the work contained in \cite{MByrd4All}).  We also believe that this 
research yields the following possibilities:

\begin{enumerate}
\item  The partial transpose condition could be used to find the set 
of separable and entangled states by finding the ranges of the angles 
for which the density matrix is positive semi-definite.
\item  The $SU(4)$ parametrization enables the calculation of the distance measure 
between density matrices and then use the minimum distance to a
completely separable matrix as a measure of separability.  Applications to 
other measures of entanglement \cite{Vedral/Plenio/Rippin/Knight} are straight-forward.
\item  One could use ranges of the angles that correspond to entangled 
states to find the ranges of the parameters in the parametrization 
in terms of the Pauli basis states by using the following 
parametrization for the density matrix
\begin{equation}
\rho = \frac{1}{4}(\Bid_{4} + a_i \sigma_i \otimes \Bid_{2} + \Bid_{2} 
\otimes b_j \tau_j 
+c_{kl} \sigma_k \otimes \tau_l).
\end{equation}
For more on this parametrization see \cite{Englert1, Englert2} and references within.
\item  Related to this last question is the question of the boundary 
between the convex set of entangled and separable states of the density
matrices.  For example one could use the explicit parametrization to calculate
specific measures of entanglement like the entanglement of formation 
for different density matrices in different regions of the 
set of density matrices and see which regions of the convex set
correspond to the greatest entanglement of formation.  Another possibility is that given
the boundary in the $\sigma, \tau$ form, we could recreate it in terms of the 
Euler angles.

\end{enumerate}
There are obviously more, but for now, it is these areas that we believe 
offer the most interest to those wishing to develop a deeper understanding 
of bipartite entanglement.  Also, since the methods here are quite general 
and rely primarily on the group theoretical techniques developed here, 
we anticipate generalizations to higher dimensional state spaces will 
be, in principle, straight-forward.


\begin{acknowledgments}
One of us (T.T.) would like to thank A.\ M.\ Kuah for his
insights into the group dynamics of this parametrization.  We would
also like to thank Dr.\ Gibbons and Dr.\ Verstraete for their input on
the $SU(4)$ covering ranges and the Peres-Horodecki eigenvalue conditions.  
Special thanks to Dr.\ Slater for his input on the two qubit density matrix
parametrization and to Dr.\ Luis Boya for his editorial assistance 
and calculations. 
\end{acknowledgments}
\pagebreak


\appendix

\section{Commutation Relations for $SU(4)$}
\label{app:crsu4}
We first note that the Gell-Mann type basis for the Lie algebra of
$SU(4)$ is given by the following set of matrices \cite{Greiner}:
\begin{equation}
\begin{array}{crcr}
\lambda_1 = \left( \begin{array}{cccc}
                     0 & 1 & 0 & 0 \\
                     1 & 0 & 0 & 0 \\
                     0 & 0 & 0 & 0 \\
                     0 & 0 & 0 & 0  \end{array} \right), &
\lambda_2 = \left( \begin{array}{crcr} 
                     0 & -i & 0 & 0 \\
                     i &  0 & 0 & 0 \\
                     0 &  0 & 0 & 0 \\
                     0 &  0 & 0 & 0  \end{array} \right), &
\lambda_3 = \left( \begin{array}{crcr} 
                     1 &  0 & 0 & 0 \\
                     0 & -1 & 0 & 0 \\
                     0 &  0 & 0 & 0 \\
                     0 &  0 & 0 & 0  \end{array} \right), \\
\lambda_4 = \left( \begin{array}{clcr} 
                     0 & 0 & 1 & 0 \\
                     0 & 0 & 0 & 0 \\
                     1 & 0 & 0 & 0 \\
                     0 & 0 & 0 & 0  \end{array} \right), &
\lambda_5 = \left( \begin{array}{crcr} 
                     0 & 0 & -i & 0 \\
                     0 & 0 &  0 & 0 \\
                     i & 0 &  0 & 0 \\
                     0 & 0 &  0 & 0 \end{array} \right), &
\lambda_6 = \left( \begin{array}{crcr} 
                     0 & 0 & 0 & 0 \\
                     0 & 0 & 1 & 0 \\
                     0 & 1 & 0 & 0 \\
                     0 & 0 & 0 & 0 \end{array} \right), \\
\lambda_7 = \left( \begin{array}{crcr} 
                     0 & 0 &  0 & 0 \\
                     0 & 0 & -i & 0 \\
                     0 & i &  0 & 0 \\
                     0 & 0 &  0 & 0 \end{array} \right), &
\lambda_8 = \frac{1}{\sqrt{3}}\left( \begin{array}{crcr} 
                     1 & 0 &  0 & 0 \\
                     0 & 1 &  0 & 0 \\
                     0 & 0 & -2 & 0 \\
                     0 & 0 &  0 & 0  \end{array} \right), &
\lambda_9 = \left( \begin{array}{crcr} 
                     0 & 0 & 0 & 1 \\
                     0 & 0 & 0 & 0 \\
                     0 & 0 & 0 & 0 \\
                     1 & 0 & 0 & 0 \end{array} \right), \\
\lambda_{10} = \left( \begin{array}{crcr} 
                     0 & 0 & 0 & -i \\
                     0 & 0 & 0 &  0 \\
                     0 & 0 & 0 &  0 \\
                     i & 0 & 0 &  0  \end{array} \right), &
\lambda_{11} = \left( \begin{array}{crcr} 
                     0 & 0 & 0 & 0 \\
                     0 & 0 & 0 & 1 \\
                     0 & 0 & 0 & 0 \\
                     0 & 1 & 0 & 0\end{array} \right), &
\lambda_{12} = \left( \begin{array}{crcr} 
                     0 & 0 & 0 &  0 \\
                     0 & 0 & 0 & -i \\
                     0 & 0 & 0 &  0 \\
                     0 & i & 0 &  0  \end{array} \right), \\
\lambda_{13} = \left( \begin{array}{crcr} 
                     0 & 0 & 0 & 0 \\
                     0 & 0 & 0 & 0 \\
                     0 & 0 & 0 & 1 \\
                     0 & 0 & 1 & 0  \end{array} \right), &
\lambda_{14} = \left( \begin{array}{crcr} 
                     0 & 0 & 0 &  0 \\
                     0 & 0 & 0 &  0 \\
                     0 & 0 & 0 & -i \\
                     0 & 0 & i &  0  \end{array} \right), &
\lambda_{15} = \frac{1}{\sqrt{6}}\left( \begin{array}{crcr}
                     1 & 0 & 0 &  0 \\
                     0 & 1 & 0 &  0 \\
                     0 & 0 & 1 &  0 \\
                     0 & 0 & 0 & -3  \end{array} \right).
\end{array}    
\end{equation}

In order to develop the Cartan decomposition of $SU(4)$ it is helpful to
look at the commutator relationships between the 15 elements of its Lie algebra.
In the following tables we list the commutator solutions of the
corresponding \textit{i}th row and \textit{j}th column Gell-Mann
matrices corresponding to the following definitions
\begin{align}
[\lambda_i , \lambda_j] =& 2if_{ijk}\lambda_k, \nonumber \\
f_{ijk} =& \frac{1}{4i}Tr[[\lambda_i , \lambda_j]\cdot \lambda_k]. \nonumber
\end{align}
\\
\\
\textit{Table 1 : $[k_1, k_2] \in L(K)$}\\
\\
(on the following page) This table corresponds to the $L(K)$ 
subset of $SU(4)$, $\{\lambda_{1},\ldots,\lambda_{8}, \lambda_{15}\}$ 
and shows that for $k_1, k_2 \in L(K)$, $[k_1, k_2] \in L(K)$.
\\
\\
\textit{Table 2 : $[p_1, p_2] \in L(K)$}\\
\\
(on the second page) This table corresponds to the $L(P)$ subset of $SU(4)$, $\{
\lambda_{9},\ldots,\lambda_{14}\}$ and shows that for $p_1, p_2
\in L(P)$, $[p_1, p_2] \in L(K)$.
\\
\\
\textit{Table 3 : $[k_1, p_2] \in L(P)$}\\
\\
(on the third page) This table corresponds to the commutator solutions for the situation
when $k_1 \in L(K)$ and $p_2 \in L(P)$, $[k_1, p_2] \in L(P)$.

\begin{turnpage}
\centering
\squeezetable

\begin{table}
\caption{$[k_1, k_2] \in L(K)$}
\begin{ruledtabular}
\begin{tabular}{|c|c|c|c|c|c|c|c|c|c|}
\hline
 & $\lambda_1$ & $\lambda_2$ & $\lambda_3$ & $\lambda_4$ & $\lambda_5$ &
 $\lambda_6$ & $\lambda_7$ & $\lambda_8$ & $\lambda_{15}$ \\ \hline
$\lambda_1$&0&$2i\lambda_3$&$-2i\lambda_2$&$i\lambda_7$&$-i\lambda_6$&$i\lambda_5$&$-i\lambda_4$&0&0 \\ \hline
$\lambda_2$&-2i$\lambda_3$&0&$2i\lambda_1$&$i\lambda_6$&$i\lambda_7$&$-i\lambda_4$&$-i\lambda_5$&0&0 \\ \hline
$\lambda_3$&$2i\lambda_2$&$-2i\lambda_1$&0&$i\lambda_5$&$-i\lambda_4$&$-i\lambda_7$&$i\lambda_6$&0&0 \\ \hline
$\lambda_4$&$-i\lambda_7$&$-i\lambda_6$&$-i\lambda_5$&0&$i(\lambda_3+\sqrt{3}\lambda_8)$&
$i\lambda_2$&$i\lambda_1$&$-i\sqrt{3}\lambda_5$&0 \\ \hline
$\lambda_5$&$i\lambda_6$&$-i\lambda_7$&$i\lambda_4$&$-i(\lambda_3+\sqrt{3}\lambda_8)$&
0&$-i\lambda_1$&$i\lambda_2$&$i\sqrt{3}\lambda_4$&0 \\ \hline
$\lambda_6$&$-i\lambda_5$&$i\lambda_4$&$i\lambda_7$&$-i\lambda_2$&$i\lambda_1$&0&$i(-\lambda_3+\sqrt{3}\lambda_8)$&$-i\sqrt{3}\lambda_7$&0 \\ \hline
$\lambda_7$&$i\lambda_4$&$i\lambda_5$&$-i\lambda_6$&$-i\lambda_1$&$-i\lambda_2$&$i(\lambda_3-\sqrt{3}\lambda_8)$&0&$i\sqrt{3}\lambda_6$&0 \\ \hline
$\lambda_8$&0&0&0&$i\sqrt{3}\lambda_5$&$-i\sqrt{3}\lambda_4$&$i\sqrt{3}\lambda_7$&$-i\sqrt{3}\lambda_6$&0&0 \\ \hline
$\lambda_{15}$&0&0&0&0&0&0&0&0&0 \\ \hline
\end{tabular}
\end{ruledtabular}
\end{table}

\begin{table}
\caption{ $[p_1, p_2] \in L(K)$}
\begin{ruledtabular}
\begin{tabular}{|c|c|c|c|c|c|c|}
\hline
& $\lambda_9$ & $\lambda_{10}$ &
$\lambda_{11}$ & $\lambda_{12}$ & $\lambda_{13}$ & $\lambda_{14}$ \\ \hline
$\lambda_9$&0&$i(\lambda_3+\frac{1}{\sqrt{3}}\lambda_8+2\sqrt{\frac{2}{3}}\lambda_{15})$&$i\lambda_2$&$i\lambda_1$&$i\lambda_5$&$i\lambda_4$ \\ \hline
$\lambda_{10}$&$-i(\lambda_3+\frac{1}{\sqrt{3}}\lambda_8+2\sqrt{\frac{2}{3}}\lambda_{15})$&0&$-i\lambda_1$&$i\lambda_2$&$-i\lambda_4$&$i\lambda_5$ \\ \hline
$\lambda_{11}$&$-i\lambda_2$&$i\lambda_1$&0&$i(-\lambda_3+\frac{1}{\sqrt{3}}\lambda_8+2\sqrt{\frac{2}{3}}\lambda_{15})$&$i\lambda_7$&$i\lambda_6$ \\ \hline
$\lambda_{12}$&$-i\lambda_1$&$-i\lambda_2$&$i(\lambda_3-\frac{1}{\sqrt{3}}\lambda_8-2\sqrt{\frac{2}{3}}\lambda_{15})$&0&$-i\lambda_6$&$i\lambda_7$ \\ \hline
$\lambda_{13}$&$-i\lambda_5$&$i\lambda_4$&$-i\lambda_7$&$i\lambda_6$&0&$2i(-\frac{1}{\sqrt{3}}\lambda_8+\sqrt{\frac{2}{3}}\lambda_{15})$ \\ \hline
$\lambda_{14}$&$-i\lambda_4$&$-i\lambda_5$&$-i\lambda_6$&$-i\lambda_7$&$2i(\frac{1}{\sqrt{3}}\lambda_8-\sqrt{\frac{2}{3}}\lambda_{15})$&0 \\ \hline
\end{tabular}
\end{ruledtabular}
\end{table}

\begin{table}
\caption{$[k_1, p_2] \in L(P)$}
\begin{ruledtabular}
\begin{tabular}{|c|c|c|c|c|c|c|c|c||c|c|c|c|c|c||c|}
\hline
& $\lambda_1$ & $\lambda_2$ & $\lambda_3$ & $\lambda_4$ & $\lambda_5$
& $\lambda_6$ & $\lambda_7$ & $\lambda_8$ & $\lambda_9$ &
$\lambda_{10}$ & $\lambda_{11}$ & $\lambda_{12}$ & $\lambda_{13}$ 
& $\lambda_{14}$ & $\lambda_{15}$ \\ \hline
$\lambda_1$ &&&&&&&&&$i\lambda_{12}$&$-i\lambda_{11}$&$i\lambda_{10}$&$-i\lambda_9$&0&0& \\ \hline
$\lambda_2$ &&&&&&&&&$i\lambda_{11}$&$i\lambda_{12}$&$-i\lambda_9$&$-i\lambda_{10}$&0&0& \\ \hline
$\lambda_3$ &&&&&&&&&$i\lambda_{10}$&$-i\lambda_9$&$-i\lambda_{12}$&$i\lambda_{11}$&0&0& \\ \hline
$\lambda_4$ &&&&&&&&&$i\lambda_{14}$&$-i\lambda_{13}$&0&0&$i\lambda_{10}$&$-i\lambda_9$& \\ \hline
$\lambda_5$ &&&&&&&&&$i\lambda_{13}$&$i\lambda_{14}$&0&0&$-i\lambda_9$&$-i\lambda_{10}$& \\ \hline
$\lambda_6$ &&&&&&&&&0&0&$i\lambda_{14}$&$-i\lambda_{13}$&$i\lambda_{12}$&$-i\lambda_{11}$& \\ \hline
$\lambda_7$ &&&&&&&&&0&0&$i\lambda_{13}$&$i\lambda_{14}$&$-i\lambda_{11}$&$-i\lambda_{12}$& \\ \hline
$\lambda_8$
&&&&&&&&&$\frac{i}{\sqrt{3}}\lambda_{10}$&$-\frac{i}{\sqrt{3}}\lambda_9$&$\frac{i}{\sqrt{3}}\lambda_{12}$&$-\frac{i}{\sqrt{3}}\lambda_{11}$&$-\frac{i}{\sqrt{3}}\lambda_{14}$&$\frac{i}{\sqrt{3}}\lambda_{13}$&
\\ \hline \hline
$\lambda_9$ &$-i\lambda_{12}$&$-i\lambda_{11}$&$-i\lambda_{10}$&$-i\lambda_{14}$&$-i\lambda_{13}$&0&0&$-\frac{i}{\sqrt{3}}\lambda_{10}$&&&&&&&$-i\sqrt{\frac{8}{3}}\lambda_{10}$ \\ \hline
$\lambda_{10}$ &$i\lambda_{11}$&$-i\lambda_{12}$&$i\lambda_9$&$i\lambda_{13}$&$-i\lambda_{14}$&0&0&$\frac{i}{\sqrt{3}}\lambda_9$&&&&&&&$i\sqrt{\frac{8}{3}}\lambda_9$ \\ \hline
$\lambda_{11}$ &$-i\lambda_{10}$&$i\lambda_9$&$i\lambda_{12}$&0&0&$-i\lambda_{14}$&$-i\lambda_{13}$&$-\frac{i}{\sqrt{3}}\lambda_{12}$&&&&&&&$-i\sqrt{\frac{8}{3}}\lambda_{12}$\\ \hline
$\lambda_{12}$ &$i\lambda_9$&$i\lambda_{10}$&$-i\lambda_{11}$&0&0&$i\lambda_{13}$&$-i\lambda_{14}$&$\frac{i}{\sqrt{3}}\lambda_{11}$&&&&&&&$i\sqrt{\frac{8}{3}}\lambda_{11}$\\ \hline
$\lambda_{13}$ &0&0&0&$-i\lambda_{10}$&$i\lambda_9$&$-i\lambda_{12}$&$i\lambda_{11}$&$\frac{i}{\sqrt{3}}\lambda_{14}$&&&&&&&$-i\sqrt{\frac{8}{3}}\lambda_{14}$\\ \hline
$\lambda_{14}$ &0&0&0&$i\lambda_9$&$i\lambda_{10}$&$i\lambda_{11}$&$i\lambda_{12}$&$-\frac{i}{\sqrt{3}}\lambda_{13}$&&&&&&&$i\sqrt{\frac{8}{3}}\lambda_{13}$\\ \hline \hline
$\lambda_{15}$
&&&&&&&&&$i\sqrt{\frac{8}{3}}\lambda_{10}$&$-i\sqrt{\frac{8}{3}}\lambda_9$&$i\sqrt{\frac{8}{3}}\lambda_{12}$&$-i\sqrt{\frac{8}{3}}\lambda_{11}$&$i\sqrt{\frac{8}{3}}\lambda_{14}$&$-i\sqrt{\frac{8}{3}}\lambda_{13}$&0\\ \hline
\end{tabular}
\end{ruledtabular}
\end{table}
\end{turnpage}

\pagebreak

\section{Invariant Volume Element Normalization Calculations}
\label{app:Haar}
Before integrating $dV_{SU(4)}$ we need some group theory.  
We begin with a digression concerning the center of a
group \cite{Artin, Scott}.  If $S$ is a subset of a group $G$, then the centralizer,
$C_{G}(S)$ of $S$ in $G$ is defined by
\begin{equation}
C(S) \equiv C_{G}(S) = \{x\in G \mid \text{ if } s\in S \text{ then } xs = sx\}.
\end{equation}
For example, if $S=\{y\}$, $C(y)$ will be used instead of $C(\{y\})$.  Next,
the centralizer of $G$ in $G$ is called the center of $G$ and is denoted by
$Z(G)$ or $Z$.  
\begin{equation}
\begin{split}
Z(G) \equiv Z&= \{z\in G \mid zx=xz  \text{ for all } x\in G\}\\
&=C_{G}(G).
\end{split}
\end{equation}
Another way of writing
this is
\begin{equation}
\begin{split}
Z(G) &= \cap \{C(x) \mid x\in G\}\\
& = \{z \mid \text{if } x\in G \text{ then } z\in C(x)\}.
\end{split}
\end{equation}
In other words, the center is the set of all elements $z$ that commutes
with all other elements in the group.  Finally, the commutator $[x,y]$ of two elements $x$ and $y$ of a group $G$ is
given by the equation
\begin{equation}
[x,y]=x^{-1}y^{-1}xy.
\end{equation}

Now what we want to find is the number of elements in the center of
$SU(N)$ for $N=2,3$, and $4$.  Begin by defining the
following
\begin{equation}
Z_{n}=\text{ cyclic group of order n }\cong \mathbb{Z}_{n} \cong Z(SU(N)).
\end{equation}
Therefore, the set of all matrices which comprise the center of $SU(N)$, $Z(SU(N))$, is congruent to $Z_{N}$ since
we know that if $G$ is a finite linear group over a field \textit{F}, then the set of matrices of
the form $\Sigma c_{g}g$, where $g \in G$ and $c_g \in F$, forms an algebra (in
fact, a ring) \cite{Scott, Sattinger}.
For example, for $SU(2)$ we would have
\begin{equation}
\begin{split}
Z_{2} =& \{x\in SU(2) \mid [x,y]\in Z_1 \text{ for all } y\in SU(2)\},\\
[x,y] =&\omega \Bid_2,\\
Z_1 =&\{\Bid_2\}.
\end{split}
\end{equation}
This would be the set of all 2 by 2 matrix elements such that the commutator
relationship would yield the identity matrix multiplied by some
non-zero coefficient.  In general this can be
written as
\begin{equation}
\begin{split}
Z_{N} =& \{x\in SU(N) \mid [x,y]\in Z_1 \text{ for all } y\in SU(N)\},\\
Z_1 =&\{\Bid_N\}.
\end{split}
\end{equation}
This is similar to the result from \cite{Artin}, that shows that the 
center of the general linear
group of real matrices, $GL_{N}(\Re)$, is the group of scalar
matrices, that is, those of the from 
$\omega \mathbb{I}$, where $\mathbb{I}$ is the identity
element of the group and $\omega$ is some multiplicative constant.  For $SU(N)$, 
$\omega \mathbb{I}$ is an $N^{th}$ root of unity.   

To begin our actual search for the normalization constant for our invariant volume 
element, we first again look at the group $SU(2)$.  For this group, every element can be
written as
\begin{equation}
\begin{pmatrix}
a & b\\
-\bar{b} & \bar{a}
\end{pmatrix}
\end{equation}
where $|a|^2 + |b|^2=1$.  Again, following \cite{Artin} we can make the
following parametrization
\begin{equation}
\begin{split}
a &= y_1 - i y_2,\\
b &= y_3 - i y_4,\\
1 &= y_{1}^2 + y_{2}^2 + y_{3}^2 + y_{4}^2.
\end{split}
\end{equation}
The elements $(1,0,0,0)$ and $(-1,0,0,0)$ are anti-podal points, 
or polar points if one pictures the
group as a three-dimensional unit sphere in a 4-dimensional space
parameterized by y, and
thus comprise the elements for the center group of $SU(2)$ (i.e. $\pm
\Bid_2$).  Therefore, the center for $SU(2)$ is comprised
of two elements.

In our parametrization, the general $SU(2)$ elements are given by
\begin{equation}
\label{su2vol}
\begin{split}
D(\mu, \nu, \xi) &= e^{i\lambda_{3}\mu}e^{i\lambda_{2}\nu}e^{i\lambda_{3}\xi},\\  
dV_{SU(2)} &= \sin(2\nu)d\mu d\nu d\xi,
\end{split}
\end{equation}
with corresponding ranges
\begin{gather}
0 \le \mu,\xi \le \pi,\\
0 \le \nu \le \frac{\pi}{2}.
\end{gather}
Integrating over the volume element $dV_{SU(2)}$ with the above
ranges yields the volume of the group $SU(2)/Z_{2}$.  In other
words, the $SU(2)$ group with its two center elements identified.  In
order to get the full volume of the $SU(2)$ group, all ones need to do
is multiply the volume of $SU(2)/Z_{2}$ by the number of removed
center elements; in this case 2.

This process can be extended to the $SU(3)$ and $SU(4)$ parametrizations.  
For $SU(3)$ \cite{MByrd3Slater1, MByrd1, MByrdp1, MByrd2}
(here recast as a component of the $SU(4)$ parametrization)
\begin{equation}
SU(3) = e^{i\lambda_3\alpha_7}e^{i\lambda_2\alpha_8}e^{i\lambda_3\alpha_9}
e^{i\lambda_{5}\alpha_{10}}D(\alpha_{11}, \alpha_{12}, \alpha_{13})
e^{i\lambda_{8}\alpha_{14}}.
\end{equation}
Now, we get an initial factor of two from the $D(\alpha_{11}, \alpha_{12}, \alpha_{13})$
component.  We shall now proves 
that we get another factor of two from the $e^{i\lambda_3\alpha_9}
e^{i\lambda_{5}\alpha_{10}}$ component as well.   

From the commutation relations of the elements of 
the Lie algebra of $SU(3)$ (see \cite{MByrd1} for details) 
we see that $\{\lambda_3, \lambda_4, \lambda_5, \lambda_8\}$ form a closed subalgebra
$SU(2)\times U(1)$.\footnote{Georgi \cite{Georgi} has stated that $\lambda_2,
\lambda_5, \text{ and } \lambda_7$ generate an $SU(2)$
subalgebra of $SU(3)$.  This fact can be seen in the commutator
relationships between these three $\lambda$ matrices contained in
\cite{MByrd1} or in Appendix \ref{app:crsu4}.} 
\begin{align}
[\lambda_3, \lambda_4]=&i\lambda_5, \nonumber \\
[\lambda_3, \lambda_5]=&-i\lambda_4, \nonumber \\
[\lambda_3, \lambda_8]=&0, \nonumber \\
[\lambda_4, \lambda_5]=&i(\lambda_3+\sqrt{3}\lambda_8), \nonumber \\
[\lambda_4, \lambda_8]=&-i\sqrt{3}\lambda_5, \nonumber \\
[\lambda_5, \lambda_8]=&i\sqrt{3}\lambda_4.
\end{align}
Observation of the four $\lambda$ matrices with respect 
to the Pauli spin matrices of $SU(2)$ shows that $\lambda_4$ is the $SU(3)$ analogue of 
$\sigma_1$, $\lambda_5$ is the $SU(3)$ analogue of $\sigma_2$ and both $\lambda_3$ and $\lambda_8$ are 
the $SU(3)$ analogues of $\sigma_3$
\begin{equation}
\begin{aligned}
\sigma_1&=\begin{pmatrix}
0 & 1 \\
1 & 0
\end{pmatrix}
\quad \Longrightarrow \quad
\lambda_4=\begin{pmatrix}
0 & 0 & 1 \\
0 & 0 & 0 \\
1 & 0 & 0 
\end{pmatrix}, \\
\sigma_2&=\begin{pmatrix}
0 & -i \\
i & 0
\end{pmatrix}
\quad \Longrightarrow \quad
\lambda_5=\begin{pmatrix}
0 & 0 & -i \\
0 & 0 & 0 \\
i & 0 & 0 
\end{pmatrix}, \\
\sigma_3&=\begin{pmatrix}
1 &  0 \\
0 & -1  
\end{pmatrix}
\quad \Longrightarrow \quad
\lambda_3=\begin{pmatrix}
1 &  0 & 0 \\
0 & -1 & 0 \\
0 & 0 & 0 
\end{pmatrix}
\text{ and }
\lambda_8=\frac{1}{\sqrt{3}}\begin{pmatrix}
1 & 0 & 0 \\
0 & 1 & 0 \\
0 & 0 & -2 
\end{pmatrix}.
\end{aligned}
\end{equation}
Thus one may use either $\{\lambda_3, \lambda_5 \}$ 
or $\{\lambda_3, \lambda_5, \lambda_8 \}$ to generate an $SU(2)$ subgroup of 
$SU(3)$.  The volume of this $SU(2)$ subgroup of $SU(3)$ must be equal to the volume of the
general $SU(2)$ group; $2\pi^2$.
If we demand that any element of the $SU(2)$ 
subgroup of $SU(3)$ have similar ranges as its $SU(2)$ 
analogue\footnote{This requires a normalization
factor of $\frac{1}{\sqrt{3}}$ on the maximal range of $\lambda_8$ that is explained
by the removal of the $Z_3$ elements of $SU(3)$.}, then a multiplicative factor of 2 is required for the 
$e^{i\lambda_3\alpha_9}e^{i\lambda_{5}\alpha_{10}}$ component.\footnote{When calculating this volume 
element, it is important to remember that the closed subalgebra being used is 
$SU(2)\times U(1)$ and therefore the integrated kernel, be it derived either from 
$e^{i\lambda_3 \alpha}e^{i\lambda_5 \beta}e^{i\lambda_3 \gamma}$ or 
$e^{i\lambda_3 \alpha}e^{i\lambda_5 \beta}e^{i\lambda_8 \gamma}$, will require contributions from both
the $SU(2)$ and $U(1)$ elements.}

Finally, $SU(3)$ has a $Z_{3}$ whose elements have the generic form:
\begin{equation}
\begin{pmatrix}
\eta_1 & 0 & 0\\
0 & \eta_2 & 0\\
0 & 0 & \eta_1^{-1}\eta_2^{-1}
\end{pmatrix},
\end{equation}
where
\begin{equation}
\eta_1^3 = \eta_2^3 = 1.
\end{equation}
Solving for $\eta_1$ and $\eta_2$ yields the following elements for $Z_{3}$
\begin{equation}
\begin{pmatrix}
                     1 & 0 & 0 \\
                     0 & 1 & 0 \\
                     0 & 0 & 1 
\end{pmatrix}, \quad
\begin{array}{cc}
-\left( \begin{array}{ccc} 
                     (-1)^\frac{1}{3} & 0 & 0 \\
                     0 & (-1)^\frac{1}{3} & 0 \\
                     0 & 0 & (-1)^\frac{1}{3} \end{array} \right),
                     \quad 
\left( \begin{array}{ccc} 
                     (-1)^\frac{2}{3} & 0 & 0 \\
                     0 & (-1)^\frac{2}{3} & 0 \\
                     0 & 0 & (-1)^\frac{2}{3}\end{array} \right)
\end{array}
\end{equation}
which are the three cube roots of unity.  Combining these $SU(3)$ center elements, a
total of three, with the 2 factors of 2 from the previous discussion, 
yields a total multiplication factor of 12.  The volume
of $SU(3)$ is then
\begin{eqnarray}
V_{SU(3)} &=& 2*2*3*V(SU(3)/Z_3) \nonumber\\
&=& \sqrt{3}\,\pi^5
\end{eqnarray}
using the ranges given above for the general $SU(2)$ elements, combined with 
$0 \le \alpha_{14} \le \frac{\pi}{\sqrt{3}}$.  Explicitly:
\begin{gather}
0 \le \alpha_7, \alpha_9, \alpha_{11}, \alpha_{13} \le \pi, \nonumber \\
0 \le \alpha_8, \alpha_{10}, \alpha_{12} \le \frac{\pi}{2}, \nonumber \\
0 \le \alpha_{14} \le \frac{\pi}{\sqrt{3}}.
\end{gather}
These are modifications of \cite{MByrd3Slater1, MByrd1, MByrdp1, MByrd2, MByrdp2} and take into account the
updated Marinov group volume values \cite{Marinov2}.

For $SU(4)$ the process is similar to that used for $SU(3)$, but now with two $SU(2)$ subgroups
to worry about.  For $SU(4)$, 
\begin{equation}
U = e^{i\lambda_3 \alpha_1}e^{i\lambda_2 \alpha_2}e^{i\lambda_3 \alpha_3}e^{i\lambda_5 \alpha_4}e^{i\lambda_3 \alpha_5}e^{i\lambda_{10} \alpha_6}[SU(3)]e^{i\lambda_{15} \alpha_{15}}.
\end{equation}
Here, the two $SU(2)$ subalgebras in $SU(4)$ that we are concerned with 
are $\{\lambda_3, \lambda_4, \lambda_5, \lambda_8, \lambda_{15}\}$ and
$\{\lambda_3, \lambda_9, \lambda_{10}, \lambda_8, \lambda_{15}\}$.
Both of these $SU(2)\times U(1)\times U(1)$ subalgebras are 
represented in the parametrization of $SU(4)$ as $SU(2)$ subgroup elements, 
$e^{i\lambda_3 \alpha_3}e^{i\lambda_5 \alpha_4}$ and 
$e^{i\lambda_3 \alpha_5}e^{i\lambda_{10} \alpha_6}$.  We can see that 
$\lambda_{10}$ 
is the $SU(4)$ analogue of $\sigma_2$\footnote{We have already discussed 
$\lambda_5$ in the previous section on $SU(3)$.} and $\lambda_{15}$ is the 
 $SU(4)$ analogue to $\sigma_3$\footnote{It is the $SU(4)$ Cartan subalgebra element.}.
The demand that all $SU(2)$ subgroups of $SU(4)$ must have a volume
equal to $2\pi^2$ is equivalent to having 
the parameters of the associated elements of the $SU(2)$ subgroup 
run through similar ranges as their $SU(2)$ analogues.\footnote{This requires a normalization
factor of $\frac{1}{\sqrt{6}}$ on the maximal range of $\lambda_{15}$ that is explained
by the removal of the $Z_4$ elements of $SU(4)$.}  
As with $SU(3)$, this restriction yields an overall multiplicative factor of 4 from 
these two elements.\footnote{When calculating these volume 
elements, it is important to remember that the closed subalgebra being used is 
$SU(2)\times U(1)\times U(1)$ and therefore, as in the $SU(3)$ case, the integrated kernels will 
require contributions from appropriate Cartan subalgebra elements.  For example, the 
$e^{i\lambda_3\alpha_3}e^{i\lambda_{5}\alpha_4}$ component is an 
$SU(2)$ sub-element of the parametrization of $SU(4)$,
but in creating its corresponding $SU(2)$ subgroup volume kernel (see the $SU(3)$ discussion), 
one must remember that it is a $SU(2)\subset SU(3) \subset SU(4)$ 
and therefore the kernel only requires 
contributions from the $\lambda_3$ and $\lambda_8$ components. On the other hand, the
$e^{i\lambda_3\alpha_5}e^{i\lambda_{10}\alpha_6}$ element corresponds to a 
$SU(2)\subset SU(4)$ and
therefore, the volume kernel will require contributions from all
three Cartan subalgebra elements of $SU(4)$.}  Recalling that the $SU(3)$ element yields a 
multiplicative factor of 12, all that remains is 
to determine the multiplicative factor equivalent to the 
identification of the $SU(4)$ center, $Z_{4}$.  

The elements of the center of $SU(4)$ are similar in form to the ones from $SU(3)$;
\begin{equation}
\begin{pmatrix}
\eta_1 & 0 & 0 & 0\\
0 & \eta_2 & 0 & 0\\
0 & 0 & \eta_3 & 0\\
0 & 0 & 0 & \eta_1^{-1}\eta_2^{-1}\eta_3^{-1}
\end{pmatrix},
\end{equation}
where
\begin{equation}
\eta_1^4 = \eta_2^4 = \eta_3^4 = 1.
\end{equation}
Solving yields the 4 roots of unity: 
$
\pm \Bid_4 \text{ and } \pm i \Bid_4,
$
where $\Bid_4$ is the 4\;x\;4 identity matrix.
So we can see that $Z_{4}$ gives another factor of 4, which, when 
combined with the factor of 4 from the two $SU(2)$ subgroups,
and the factor of 12 from the $SU(3)$ elements, gives a total 
multiplicative factor of 192.  Integration of the volume 
element given in equation \eqref{dvsu4} with the following ranges
\begin{gather}
0 \le \alpha_1, \alpha_3, \alpha_5, 
\alpha_7, \alpha_9, \alpha_{11}, \alpha_{13} \le \pi, \nonumber \\
0 \le \alpha_2, \alpha_4, \alpha_6,
\alpha_8, \alpha_{10}, \alpha_{12} \le \frac{\pi}{2}, \nonumber \\
0 \le \alpha_{14} \le \frac{\pi}{\sqrt{3}}, \nonumber \\
0 \le \alpha_{15} \le \frac{\pi}{\sqrt{6}},
\end{gather}
gives
\begin{eqnarray}
V_{SU(4)} &=& 2*2*2*2*3*4*V(SU(4)/Z_4)\nonumber\\
&=&\frac{\sqrt{2}\,\pi^9}{3}.
\end{eqnarray}  
This calculated volume for $SU(4)$ agrees  
with that from Marinov \cite{Marinov2}.
\pagebreak


\section{Modified Parameter Ranges for Group Covering}
\label{app:paramranges}
In order to be complete, we list the modifications to the ranges given
in Appendix \ref{app:Haar} that affect a covering of $SU(2)$, $SU(3)$, and $SU(4)$
without jeopardizing the calculated group volumes.

To begin, in our parametrization, the general $SU(2)$ elements are given by
\begin{equation}
\begin{split}
D(\mu, \nu, \xi) &= e^{i\lambda_{3}\mu}e^{i\lambda_{2}\nu}e^{i\lambda_{3}\xi},\\  
dV_{SU(2)} &= \sin(2\nu)d\mu d\nu d\xi,
\end{split}
\end{equation}
with the corresponding ranges for the volume of $SU(2)/Z_2$ given as
\begin{gather}
0 \le \mu,\xi \le \pi,\nonumber \\
0 \le \nu \le \frac{\pi}{2}.
\end{gather}
In order to generate a covering of $SU(2)$, the $\xi$ parameter must be
modified to take into account the uniqueness of the two central group
elements, $\pm \Bid_2$, under spinor transformations.\footnote{For 
specific examples of this, see either \cite{Biedenharn} or \cite{Ryder}.} 
This modification is straightforward enough; $\xi$'s range is multiplied by the number
of central group elements in $SU(2)$.  The new ranges are thus 
\begin{gather}
0 \le \mu \le \pi,\nonumber \\
0 \le \nu \le \frac{\pi}{2},\nonumber \\
0 \le \xi \le 2\pi.
\end{gather}
These ranges yield both a covering of $SU(2)$, as well as the correct
group volume for $SU(2)$.\footnote{One may interchange $\mu$ and $\xi$'s 
ranges without altering either 
the volume calculation, or the final orientation of a two-vector under 
operation by $D$.  This interchange
is beneficial when looking at Euler parametrizations beyond $SU(2)$.}  

For $SU(3)$, here given as a component of the $SU(4)$ parametrization,
we know we have two $SU(2)$ components (from Appendix \ref{app:Haar}),
\begin{equation}
SU(3) = e^{i\lambda_3\alpha_7}e^{i\lambda_2\alpha_8}e^{i\lambda_3\alpha_9}
e^{i\lambda_{5}\alpha_{10}}D(\alpha_{11}, \alpha_{12}, \alpha_{13})
e^{i\lambda_{8}\alpha_{14}}.
\end{equation}
Therefore the ranges of $\alpha_9$ and $\alpha_{13}$ should be modified
just as $\xi$'s was done in the previous discussion for $SU(2)$.  Remembering the discussion in
Appendix \ref{app:Haar} concerning the central group of $SU(3)$, we can deduce that
$\alpha_{14}$'s ranges should be multiplied by a factor of 3.  This
yields the following, corrected, ranges for $SU(3)$\footnote{Earlier
representations of these ranges for $SU(3)$, for example in
\cite{MByrd3Slater1, MByrd1, MByrdp1, MByrd2, Gibbons, MByrdp2}, were
incorrect in that they failed to take into account the updated $SU(N)$ volume formula in \cite{Marinov2}.}
\begin{gather}
0 \le \alpha_7,\alpha_{11} \le \pi,\nonumber \\
0 \le \alpha_8,\alpha_{10},\alpha_{12} \le \frac{\pi}{2},\nonumber \\
0 \le \alpha_9,\alpha_{13} \le 2\pi,\nonumber \\
0 \le \alpha_{14} \le \sqrt{3}\pi.
\end{gather}
These ranges yield both a covering of $SU(3)$, as well as the correct
group volume for $SU(3)$.

For $SU(4)$, we have two $SU(2)$ subgroup components
\begin{equation}
SU(4) = e^{i\lambda_3 \alpha_1}e^{i\lambda_2 \alpha_2}e^{i\lambda_3 \alpha_3}e^{i\lambda_5 \alpha_4}e^{i\lambda_3 \alpha_5}e^{i\lambda_{10} \alpha_6}[SU(3)]e^{i\lambda_{15} \alpha_{15}}.
\end{equation}
As with the $SU(2)$ subgroup ranges in $SU(3)$, the ranges for $\alpha_3$ and $\alpha_5$ 
each get multiplied by 2 and
$\alpha_{15}$'s ranges get multiplied by 4 (the number of $SU(4)$
central group elements).  The remaining ranges are either held
the same, or modified in the case of the $SU(3)$ element;
\begin{gather}
0 \le \alpha_1,\alpha_7,\alpha_{11} \le \pi,\nonumber \\
0 \le \alpha_2,\alpha_4,\alpha_6,\alpha_8,\alpha_{10},\alpha_{12}
\le \frac{\pi}{2},\nonumber \\
0 \le \alpha_3,\alpha_5,\alpha_9,\alpha_{13} \le 2\pi,\nonumber \\
0 \le \alpha_{14} \le \sqrt{3}\pi,\nonumber \\
0 \le \alpha_{15} \le 2\sqrt{\frac{2}{3}}\pi.
\end{gather}
These ranges yield both a covering of $SU(4)$, as well as the correct
group volume for $SU(4)$.

In general we can see that by looking at 
$SU(N)/Z_N$ not only can we arrive at a parametrization of $SU(N)$ with a logically derivable set
of ranges that gives the correct group volume, but we can also show 
how those ranges can be logically modified to cover the entire group as well without any
arbitrariness in assigning values to the parameters.  It is this work that will be the 
subject of a future paper.  
\pagebreak




\end{document}